
\input phyzzx

 \def\leftrightarrowfill{$\mathord-\mkern-6mu%
   \cleaders\hbox{$\mkern-2mu\mathord-\mkern-2mu$}\hfill
   \mkern-6mu\mathord\leftrightarrow$}
 \def\overlrarrow#1{\vbox{\ialign{##\crcr
       \leftrightarrowfill\crcr\noalign{\kern-1pt\nointerlineskip}
       $\hfil\displaystyle{#1}\hfil$\crcr}}}

\def\dlrmd{\overlrarrow{\partial_{\mu}}}
\def\dlrnd{\overlrarrow{\partial_{\nu}}}

\def\smij{\bar{\psi}_i  \dlrmd \gamma_5 \psi_i \, \bar{\psi}_j
  \gamma^{\mu} \psi_j}
\def\emij{\bar{\psi}_i  \dlrnd \gamma_5 \psi_i \,
  \partial_{\mu} F^{\mu \nu}}

\def\Cs{$^{133}$Cs}
\def\Tl{$^{205}$Tl}
\def\Hg{$^{199}$Hg}
\def\Xe{$^{129}$Xe}
\def\TlF{$^{205}$TlF}

\def\ls#1{_{\lower1.5pt\hbox{$\scriptstyle #1$}}}


\frontpagetrue


\let\picnaturalsize=N
\def\picsize{1.0in}
\def\picfilename{scipp_tree.eps}

\let\nopictures=Y

\ifx\nopictures Y\else{\ifx\epsfloaded Y\else\input epsf \fi
\let\epsfloaded=Y
{\line{\hbox{\ifx\picnaturalsize N\epsfxsize \picsize\fi
{\epsfbox{\picfilename}}}\hfill\vbox{


\hbox{SCIPP 93/45}
\hbox{December, 1993}
\vskip1.2in
}
}}}\fi


\def\SCIPP{\centerline {\it Santa Cruz Institute for Particle Physics}
  \centerline{\it University of California, Santa Cruz, CA 95064}}
\overfullrule 0pt

\pubtype{ T}     

\hfill\hbox{SCIPP 93/45}

\vskip1.in

\title{{Electromagnetic Contributions to the Schiff Moment}
\foot{Supported in part by the U.S. Department of Energy
and the Texas National Research Laboratory Commission under grant
numbers RGFY 93-263 and RGFY 93-330.}}
\author{Scott Thomas
}
\SCIPP
\vskip1cm
\vbox{
\centerline{\bf Abstract}

\parskip 0pt
\parindent 25pt
\overfullrule=0pt
\baselineskip=19pt
\tolerance 3500
\endpage
\pagenumber=1
\singlespace
\bigskip

The Schiff moment, $\smij$, is a parity and time reversal
violating fermion-fermion coupling. The nucleus-electron Schiff moment
generically gives the most important contribution to the electric
dipole moments of atoms and molecules with zero net intrinsic
electronic spin and nuclear spin ${1 \over 2}$.  Here, the electromagnetic
contribution to the Schiff moment, $\emij$, is considered.
For a nucleon, the leading chirally violating contribution to
this interaction is calculable in the chiral limit
in terms of the parity and
time reversal violating pion-nucleon coupling.
For the Schiff moment of heavy nuclei, this chiral
contribution is somewhat smaller than
the finite size effect discussed previously in the literature.
}

\vfill
\submit{Physical Review D}
\vfill

\singlespace
\bigskip


\baselineskip=20pt

\REF\schiff{L. Schiff, Phys. Rev. {\bf 132}, 2194 (1963).}

\REF\susycp{W. Fischler, S. Paban, and S. Thomas, Phys. Lett. B
{\bf 289}, 373 (1992).}

\REF\khrip{O. Sushkov, V. Flambaum, and I. Khriplovich,
Zh. Eksp. Teor. Fiz. {\bf 87}, 1521 (1984)
[Sov. Phys. JETP {\bf 60}, 873 (1984)].}

\REF\com{I. Khriplovich, Comments At. Mol. Phys. {\bf 22},
295 (1989).}

\REF\book{I. Khriplovich, {\it Parity Nonconservation in Atomic
Phenomena} (Gordon and Breach, London, 1991).}

Experimental searches for electric dipole moments (EDMs)
provide low energy probes for P and T violating physics at and
beyond the electroweak symmetry breaking scale.
The current experimental limits on the EDMs of atoms, molecules,
and the neutron already give bounds on certain T violating
extensions of the standard model.
The sensitivity of atomic and molecular EDMs to
microscopic P and T violation depends on the net intrinsic
electronic spin.
Atoms with an unpaired electron, such as \Cs~and \Tl,
are sensitive mainly to the electron EDM.
Atoms with paired electrons, such as \Hg~ and \Xe, or
the molecule \TlF, are sensitive mainly to nuclear effects.
The effect of the nuclear EDM is highly suppressed due
to Schiff's theorem.\refmark{\schiff}
Higher electromagnetic moments are not affected by Schiff's
theorem.
This allows the magnetic quadrupole moment to contribute for
nuclei with $J \geq$ 1, where $J$ is the nuclear spin.
For $J={1 \over 2}$ however, the dominant nuclear contribution
to the atomic or molecular EDM comes from a local (on the atomic
scale) coupling between the nucleus and electrons known as
the Schiff moment.

The Schiff moment (SM) coupling two spin ${1 \over 2}$ Dirac
fermions arises from the operators
$$
S_1~ \smij
\eqn\sm
$$
$$
- S_2~ \bar{\psi}_i \sigma^{\mu \nu} i \gamma_5 \psi_i ~
 \partial_{\nu} ( \bar{\psi}_j \gamma_{\mu} \psi_j)
\eqn\smb
$$
These operators are equivalent on shell and can therefore
be related using equations of motion.
In what follows only \sm~will be kept
explicitly.
The contribution of the nucleus-electron SM
to an atomic EDM can
be estimated just on dimensional grounds to be
$d_a \sim eS~ Z^2 \alpha m_e^2$, where $m_e$ is
the electron mass.\refmark{\khrip- \book}
Similarly, a molecular EDM can be estimated to be
$d_m \sim eS~ Z^2 \alpha m_e m_N$, where $m_N$ is the
nucleus mass.\refmark{\khrip - \book}

The magnitude of $S$ depends on the origin of the microscopic
P and T violation and the scale at which the effective operator
\sm~is generated.
One contribution comes from the neutral current component of
the weak electric dipole moment (WEDM),
$-d_{\rm w}~ {1 \over 2}
\bar{\psi}_i \sigma^{\mu \nu} i \gamma_5 \psi_i
Z_{\mu \nu}$, where $Z_{\mu \nu}$ is the $Z$ boson field strength.
Such an operator would be generated, for example,
at one loop in a multi-Higgs model of T violation.
For light fermions the chirally violating WEDM
is effectively dimension six, being suppressed by two powers of
the heavy scale associated with T violation.
Tree level $Z$ exchange, with the WEDM,
then gives a SM \sm~with
$S= g_{\rm v} d_{\rm w} / m_{\rm z}^2$, where
$g_{\rm v}$ is the neutral current vector coupling of
the fermion $j$.\refmark{\susycp}
A SM can also arise directly
at the heavy scale.
For example, in the supersymmetric standard model box
diagrams involving gauginos and the scalar partners of
the external fermions give
$S \sim \sin \phi ~\alpha^2 m_i / M^4_{\rm SUSY}
$, where $\sin \phi$ is some combination of T violating
phases.\refmark{\susycp}
The nucleus-electron SM arising from
quark-electron moments of the type discussed above
are suppressed by four powers of a heavy mass.
The resulting atomic or molecular EDM is therefore
less important than that arising from, for example, the light
quark EDM or chromo-electric dipole moment (CEDM), which
are suppressed by only two powers of a heavy mass.

A SM may also arise from the electromagnetic interaction
$$
S^{\prime}~\emij
\eqn\sme
$$
Using the equation of motion
$ \partial_{\mu} F^{\mu \nu} = e Q_j ~
   \bar{\psi}_j \gamma^{\nu} \psi_j $ gives the
operator \sm~with
$S=eQ_jS^{\prime}$.
Diagrammatically, \sm~arises from the coupling
of the electromagnetic current to \sme~through
tree level photon exchange.
The $q^2$ dependence in \sme~is canceled by the
photon propagator.
In order to make explicit the origin of the operator \sme~consider
the matrix element of the electromagnetic current for particle
$i$.
The most general P and T odd matrix element of $j_{\mu}$
on two single particle Dirac states
can
be written
$$
\langle p^{\prime}, s^{\prime} | j^{\mu} | p , s \rangle =
D(q^2) ~
\bar{u}(p^{\prime}, s^{\prime}) \sigma^{\mu \nu} \gamma_5
  q_{\nu} u(p,s)
\eqn\mat
$$
where $q_{\nu} = (p - p^{\prime})_{\nu}$, and
$D(q^2)$ is a momentum dependent form factor.
Expanding about $q^2 =0$, the
constant part of $D(q^2)$ is just the
electric dipole moment, $d$,
$$
d = D(0)
\eqn\edm
$$
The $q^2$ dependent piece of $D(q^2)$ is reproduced for on shell
fermions by the operator \sme~with
$$
S^{\prime} = {d \over dq^2} D(0)
\eqn\smat
$$
As long as $D(q^2)$ is not constant, any process which
produces an EDM also produces a SM.
\foot{It is worth noting that Eqs. \mat~and \smat~show that the
electromagnetic contribution to the SM can be thought of as
the P and T odd analog of the charge radius or
electromagnetic anapole moment.}
In particular, a quark-electron electromagnetic
SM can be generated at the
heavy scale associated with T violation by the same
processes responsible for a quark EDM.
However,
this will necessarily be suppressed by four powers of the heavy
scale, just as the previous contributions.

\REF\dn{R. Crewther, P. Di Vecchia, G. Veneziano, and E. Witten,
Phys. Lett. B {\bf 88}, 123 (1979); Phys. Lett. B {\bf 91}, 487(E)
(1980).}

\FIG\figloop{Nonanalytic contribution to the nucleon electromagnetic
Schiff moment from the coupling $\bar{g} \bar{N} \pi N$.
Other graphs related by gauge invariance are not shown.
The graph with the photon attached to the nucleon is smaller by
${\cal O}(m_{\pi}^2 / m_n^2)$.}

More important are contributions to the nucleus-electron
SM arising at the nuclear scale.
First consider the nucleon-electron SM.
In the chiral limit, effective operators involving nucleons
are typically dominated by nonanalytic contributions
arising from integrating out pions.
As an example of the nonanalytic contribution to the nucleon
SM consider the chirally violating P and T odd pion-nucleon coupling
$$
\bar{g} \bar{N} \pi N
\eqn\ptodd
$$
where $\pi = \tau^a \pi^a$.
This coupling could arise from a finite QCD vacuum angle
or a light quark CEDM.\refmark{\susycp}
The nonanalytic contribution comes from the same
graphs which give a nucleon EDM (see fig. \figloop).\refmark{\dn}
This graph for the SM is divergent in the
infrared, cutoff by the pion mass.
A straightforward calculation gives
$$
S^{\prime} = { e \bar{g} g_A \over 48 \pi^2 f_{\pi} m_{\pi}^2 }
\eqn\sloop
$$
where $g_A \simeq 1.26$ is the usual pion-nucleon
coupling, and $f_{\pi} \simeq$ 93 MeV is
the pion decay constant.
Notice that since $\bar{g}$ scales as $m_{\pi}^2$,
$S^{\prime}$ is a constant in the chiral limit.
This is in contrast to the nucleon EDM from \ptodd~which
scales as $m_{\pi}^2 \ln m_{\pi}^2$ in the chiral
limit.\refmark{\dn}
The SM of individual nucleons will contribute incoherently
to the nucleus SM.
This nonanalytic contribution to the SM of heavy
spin ${1 \over 2}$ nuclei will therefore be
equal to the nucleon SM \sloop.

\REF\nuc{V. Flambaum, I. Khriplovich, and O. Sushkov,
Phys. Lett. B {\bf 162}, 213 (1985);
Nucl. Phys. A {\bf 449}, 750 (1986).}

In addition to the incoherent contribution to the nucleus-electron
SM, there are coherent contributions due to the finite
size of the nucleus.
The magnitude of the finite size effects may be estimated with
a simple model due to Sushkov, Flambaum, and
Khriplovich.\refmark{\khrip - \book,\nuc}
The model assumes a nucleus with a single unpaired
valence proton.
In the nonrelativistic limit \ptodd~leads to a coupling
of the valence proton to the nuclear core of
$$
{\bar{g} g_A \over f_{\pi} m_{\pi}^2 }
{}~\vec{\sigma} \cdot \vec{\nabla} \rho
\eqn\vcoup
$$
where $\rho$ is the core density.
To model the effect of this interaction
the nuclear potential, $U$, is assumed to be proportional
to $\rho$, i.e.
$U = \rho (U_o / \rho_o)$, where
$\rho_o \sim \tilde{m}^3$ and $U_o \sim \tilde{m}^3 / 4 \pi f_{\pi}^2$
are the density and potential deep in the core,
and $\tilde{m}$ is some nuclear mass parameter
characterizing the repulsive part of the nuclear potential
($\tilde{m}$ is independent of the chiral limit).
Under this assumption the interaction \vcoup~leads to a constant
shift of the valence proton wave function given by
$$
\vec{\lambda} \simeq {4 \pi \bar{g} g_A f_{\pi} \over
      m_{\pi}^2} ~\vec{\sigma}
\eqn\shift
$$
This constant shift leads to P and T odd interactions of the
nucleus with the electromagnetic field.
For spin ${1 \over 2}$
these are contained in the form factor $D(q^2)$.
Just on dimensional grounds the electromagnetic
SM resulting from the shift
$\vec{\lambda}$ is\refmark{\khrip,\nuc}
$$
S^{\prime} \sim {e \over 4 \pi} \lambda R^2
\sim  {e \bar{g} g_A f_{\pi} A^{2/3} \over
m_{\pi}^2 \tilde{m}^2 }
\eqn\scoh
$$
where $R \sim A^{1/3} \tilde{m}^{-1}$ is the
rms radius of the valence wave function, and
$A$ the atomic number.
The finite size effect is essentially coherent over the entire
nucleus, being proportional to the square of the valence nucleon
wave function radius.
It is larger than the incoherent loop contribution \sloop~by
${\cal O}((4 \pi f_{\pi} / \tilde{m})^2 A^{2/3})$, where
$(4 \pi)^2$ counts the loop factor.


In conclusion, any P and T violating microscopic physics
which generates EDMs also generally gives rise to
an electromagnetic SM.
The nucleus-electron SM represents the most important
nuclear contribution to the EDMs of atoms and
molecules with paired electrons and
nuclear spin ${1 \over 2}$.
This moment arises predominantly from indirect
effects at the nuclear
scale rather than directly from the microscopic P and
T violating scale.
The chirally violating contribution to the nucleon SM
is nonanalytic and calculable in the chiral limit.
Coherent finite size effects however dominate the
SM of heavy nuclei.

\refout
\figout

\end